\documentclass[twocolumn]{aastex631}
\usepackage{graphicx}
\usepackage{float}
\usepackage{amsmath}
\usepackage{amssymb}
\usepackage{subfigure}
\usepackage{cases}
\usepackage{mathrsfs}
\usepackage[flushleft]{threeparttable}
\usepackage{amssymb}
\usepackage{pifont}
\usepackage{epstopdf}
\usepackage{xcolor}
\usepackage[all]{hypcap}
\usepackage{mwe,tikz}
\usepackage{bm}
\usepackage{tabularx}
\usepackage{multirow}
\usepackage{longtable}
\usepackage{xspace}
\usepackage{rotating}
\usepackage{CJK}
\usepackage{upgreek}
\usepackage{booktabs}
\usepackage{textcomp}
\usepackage{makecell}
\usepackage{enumitem}
\usepackage{csquotes}
\usepackage{soul}

\newcommand{\e}{\text{e}}

\newcommand{\p}{\partial}

\begin{document}
 \title{A Comptonized Fireball Bubble Fits the Second Extragalactic Magnetar Giant Flare GRB 231115A}
\correspondingauthor{Bin-Bin Zhang}
\email{bbzhang@nju.edu.cn}
\author[0000-0002-5596-5059]{Yi-Han Iris Yin}
\affiliation{School of Astronomy and Space Science, Nanjing University, Nanjing 210093, China}
\affiliation{Key Laboratory of Modern Astronomy and Astrophysics (Nanjing University), Ministry of Education, China}
\author[0000-0001-6869-2996]{Zhao Joseph Zhang}
\affiliation{Theoretical Astrophysics, Department of Earth and Space Science, Osaka University, 1-1 Machikaneyama, Toyonaka, Osaka 560-0043, Japan}
\author[0000-0002-5485-5042]{Jun Yang}
\affiliation{School of Astronomy and Space Science, Nanjing University, Nanjing 210093, China}
\affiliation{Key Laboratory of Modern Astronomy and Astrophysics (Nanjing University), Ministry of Education, China}
\author[0009-0009-2083-1999]{Run-Chao Chen}
\affiliation{School of Astronomy and Space Science, Nanjing University, Nanjing 210093, China}
\affiliation{Key Laboratory of Modern Astronomy and Astrophysics (Nanjing University), Ministry of Education, China}
\author[0000-0003-3071-9123]{Umer Rehman}
\affiliation{School of Astronomy and Space Science, Nanjing University, Nanjing 210093, China}
\affiliation{Key Laboratory of Modern Astronomy and Astrophysics (Nanjing University), Ministry of Education, China}
\affiliation{Department of Physics, Air University, E-9 Sector PAF complex 44000 Islamabad, Pakistan}
\author[0000-0003-1471-2693]{Varun}
\affiliation{School of Astronomy and Space Science, Nanjing University, Nanjing 210093, China}
\affiliation{Key Laboratory of Modern Astronomy and Astrophysics (Nanjing University), Ministry of Education, China}
\author[0000-0003-4111-5958]{Bin-Bin Zhang}
\affiliation{School of Astronomy and Space Science, Nanjing University, Nanjing 210093, China}
\affiliation{Key Laboratory of Modern Astronomy and Astrophysics (Nanjing University), Ministry of Education, China}
\affiliation{Purple Mountain Observatory, Chinese Academy of Sciences, Nanjing, 210023, China}
\begin{abstract}
Magnetar giant flares (MGFs), originating from noncatastrophic magnetars, share noteworthy similarities with some short gamma-ray bursts (GRBs). However, understanding their detailed origin and radiation mechanisms remains challenging due to limited observations. The discovery of MGF GRB 231115A, the second extragalactic MGF located in the Cigar galaxy at a luminosity distance of $\sim 3.5$ Mpc, offers yet another significant opportunity for gaining insights into the aforementioned topics. This Letter explores its temporal properties and conducts a comprehensive analysis of both the time-integrated and time-resolved spectra through empirical and physical model fitting. Our results reveal certain properties of GRB 231115A that bear resemblances to GRB 200415A. We employ a Comptonized fireball bubble model, in which the Compton cloud, formed by the magnetar wind with high density $e^{\pm}$, undergoes Compton scattering and inverse Compton scattering, resulting in reshaped thermal spectra from the expanding fireball at the photosphere radius. This leads to dynamic shifts in dominant emission features over time. Our model successfully fits the observed data, providing a constrained physical picture, such as a trapped fireball with a radius of $\sim 1.95 \times 10^{5}$ cm and a high local magnetic field of $2.5\times 10^{16}$ G. The derived peak energy and isotropic energy of the event further confirm the burst's MGF origin and its contribution to the MGF-GRB sample. We also discuss prospects for further gravitational wave detection associated with MGFs, given their high-event-rate density ($\sim 8\times 10^5\ \rm Gpc^{-3}\ yr^{-1}$) and ultrahigh local magnetic field.
\end{abstract}
\section{Introduction}
\label{sec:introduction}
Magnetar giant flares (MGFs) are rare and exceptionally powerful transient phenomena originating from noncatastrophic magnetars. They exhibit light-curve structures consisting of an initial spike lasting tenths of a second, followed by a much dimmer pulsating tail modulated by the magnetar's spin. The tail is only visible within close proximity, e.g., our Galaxy \citep{1979Natur.282..587M,1983A&A...126..400B, 1999Natur.397...41H, 2005Natur.434.1098H, Israel_2005,Strohmayer_2005,10.1111/j.1745-3933.2006.00155.x}. Despite the typical temporal characteristics extracted from the observed Galactic MGF events, all of them are saturated by the gamma-ray detectors due to the overwhelming surge in photon numbers \citep{1984Natur.307...41G,2002MmSAI..73..554F,Yamazaki_2006}. Consequently, obtaining both well-featured temporal data and the accurate spectral data of MGF has been unattainable, until the first spectrally confirmed MGF GRB 200415A \citep{Yang_2020,10.1093/mnras/stad443}. This event suggests that MGFs in nearby galaxies could produce short gamma-ray bust (GRB)-like events and contribute to at least a subsample of the observed short gamma-ray bursts \citep{1986Natur.322..152L, 1987ApJ...320L.105A, 2001AIPC..586..495D, 2005Natur.434.1098H, 10.1111/j.1745-3933.2005.00062.x}.
Moreover, a comprehensive analysis and understanding of the MGF spectrum is of great importance in order to identify potential misclassified MGFs in short GRB population, which necessitates a validated and fittable MGF model that effectively accounts for the underlying cause and radiation mechanism.

To understand the mechanism of an MGF, various models involving either internal \citep{1983ApJ...264..642P, 1983ApJ...264..635P} or external \citep{1985JFM...159..359M, Thompson_2001} factors have been proposed. Radiation transfer models, such as the ``trapped fireball" and ``magnetar relativistic wind," have significantly contributed to providing clarity on the aspect of radiation mechanisms \citep{10.1093/mnras/275.2.255}. Subsequently, a composite model involving both these components was further proposed and developed in the form of the ``Comptonized fireball" \citep{10.1093/mnras/stad443}. In this model, photons from the fireball are upscattered and downscattered by the dense $e^{\pm}$ pairs at the photosphere radius, producing a multicomponent thermal-like spectrum. This physically derived model has been used and successfully fitted to the spectra of the first extragalactic MGF, GRB 200415A \citep{Zhang_2020, Burns_2021, Yang_2020, 2021Natur.589..211S}, offering a method to explore MGF spectral data and gain physical insights into MGF GRBs.

Recently, yet another MGF GRB, 231115A, was detected and initially classified as a short GRB-like event \citep{2023GCN.35035....1F}. Its MGF origin was later inferred by the positional consistency with the nearby galaxy M82 \citep[aka Cigar galaxy;][]{2023GCN.35038....1B, 2023GCN.35037....1M} and subsequent empirical spectral analysis \citep{2023GCN.35062....1F, 2023GCN.35059....1M, 2023arXiv231202848W}. This event provides us with an additional opportunity to conduct an in-depth study of MGF temporal and spectral properties using the modified Comptonized fireball model and state-of-the-art fitting tools. Such an approach will allow us to directly check if the observed properties can be fitted to the physical model, thus more directly revealing the physical origins. In this Letter, we first provide details of data reduction and analysis in section \ref{sec:temporal}. Empirical and physical model fitting are presented in section \ref{sec:spectral}. Finally, we summarize and discuss our results in section \ref{sec:sum}.
\begin{table}
\centering
\caption{Summary of the observed properties of MGF GRB 231115A.}
\label{tab:summary}
\begin{tabular}{ll}
\hline
\hline
Observed Properties & GRB 231115A \\
\hline
$T_{\rm 90}$ ($\rm ms$) & $55.90_{-1.91}^{+3.43}$ \\
Total spanning time ($\rm ms$) & $\sim\ 79$ \\
Minimum variability timescale ($\rm ms$) & $\sim\ 13.95$ \\
Spectral index $\alpha$ (CPL) & ${0.16}_{-0.19}^{+0.21}$ \\
Peak energy ($\rm keV$) (CPL) & ${605.54}_{-67.84}^{+84.72}$ \\
Peak energy ($\rm keV$) & $610.07_{-38.62}^{+110.68}$ \\
Peak flux ($\rm erg\,cm^{-2}\,s^{-1}$) & $2.02_{-0.28}^{+0.16}\times10^{-5}$ \\
Total fluence ($\rm erg\,cm^{-2}$) & $6.36_{-0.43}^{+0.48}\times10^{-7}$ \\
Peak luminosity ($\rm erg\,s^{-1}$) & $2.95_{-0.41}^{+0.23}\times10^{46}$ \\
Isotropic energy ($\rm erg$) & $9.32_{-0.63}^{+0.70}\times10^{44}$ \\
\hline
\multirow{2}{*}{Possible host galaxy} & Cigar galaxy \\
& (NGC 3034) \\
Distance (Mpc) & 3.5 \\
\hline
Event rate density ($\rm Gpc^{-3}\ yr^{-1}$) & $\sim 8\times 10^5$ \\
\hline
\hline
\end{tabular}
\end{table}
\section{Data Reduction and Analysis}
\label{sec:temporal}
At 15:36:21.201 UT on 2023 November 15 (denoted as $T_0$), the Fermi Gamma-ray Burst Monitor \citep[GBM;][]{Meegan_2009} detected the MGF GRB 231115A \citep{2023GCN.35044....1D, 2023GCN.35035....1F}. Almost immediately, INTEGRAL \citep{2003A&A...411L...1W} was also triggered by the event \citep{2023GCN.35036....1D}. Subsequently, the positional data indicated alignment with the nearby galaxy M82, situated at luminosity distance of $\sim$ 3.5 Mpc \citep{2023GCN.35038....1B, 2023GCN.35037....1M}. We retrieved the time-tagged event dataset covering the time range of MGF GRB 231115A from the Fermi/GBM public data archive\footnote{\url{https://heasarc.gsfc.nasa.gov/FTP/fermi/data/gbm/daily/}}. Among all 12 sodium iodide (NaI) detectors, n6, n7 and n8 were selected with the smallest viewing angles with respect to the GRB source direction. Additionally, for temporal and spectral analysis, we included the brightest bismuth germanium oxide detector, b1.

\begin{figure}
 \centering
 \includegraphics[width = 0.45\textwidth]{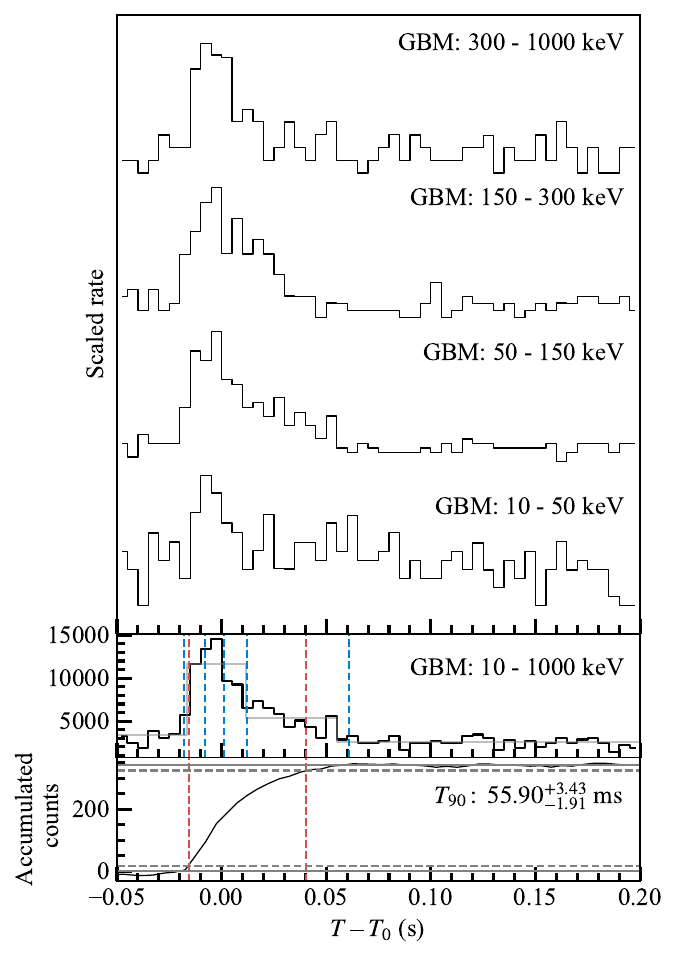}
 \caption{Multiwavelength light curves of MGF GRB 231115A (top panel). The bin size is set to 5 ms for all light curves. The bottom two panels show the light curve of Fermi/GBM in the energy range of 10-1000 keV and the accumulated counts. The red dashed vertical lines represent the $T_{\rm 90}$ interval. The gray curve corresponds to the derived Bayesian blocks from time-tagged event data. The blue dashed vertical lines mark four time slices (as listed in Table \ref{tab:specfit}) for spectral analysis.}
 \label{fig: lc_t90}
\end{figure}

We processed the Fermi/GBM data following the standard procedures described in \cite{Zhang_2011} and \cite{2022Natur.612..232Y}. Figure \ref{fig: lc_t90} demonstrates that the event lasts for $\sim$ 79 ms after $T_0$, exhibiting consistent pulse profiles in different energy ranges. Upon analyzing multiwavelength light-curve pairs in the upper four panels of Figure \ref{fig: lc_t90}, we derived spectral lags between the lowest energy band (10-50 keV) and higher energy bands (50-150, 150-300, 300-1000 keV), revealing tiny values of $0.60_{-4.30}^{+4.10}$ ms, $2.20_{-7.30}^{+4.90}$ ms and $0.50_{-2.70}^{+3.30}$ ms, respectively. We further extracted the $T_{90}$ interval of $55.90_{-1.91}^{+3.43}$ ms in the standard energy range of 10-1000 keV, as is depicted in the lower two panels of Figure \ref{fig: lc_t90}. Following the same energy range, two Bayesian blocks were recognized by implementing Bayesian method \citep{Scargle_2013} on the time-tagged event data. Half of the minimum bin size of these blocks, 13.95 ms, is regarded as the minimum variability timescale of this event. Those temporal properties fall within the expected range for a short GRB, similar to GRB 200415A \citep{Yang_2020}.

Given the consecutive detection of MGF GRBs 200415A \citep{Yang_2020, Zhang_2020, Burns_2021, 2021Natur.589..211S} and 231115A \citep{2023GCN.35062....1F, 2023arXiv231214645M, 2023GCN.35059....1M, 2023arXiv231202848W}, we calculated the event rate density ($\rho$) for extragalactic MGFs using the formula $\frac{\Omega T}{4\pi}\rho V_{\rm max}=N=2$, where $\Omega \sim 8$ sr, and $T \sim 7.5$ yr are associated with the Fermi/GBM field of view and effective operational time. We considered $D_{\rm max} \sim$ 5 Mpc from \citet{Burns_2021} as the maximum distance for detecting such an event, from which we obtained the maximum volume $V_{\rm max}$. The derived event rate density ($\rho$) is approximately $8 \times 10^{5}\ {\rm Gpc^{-3}\ yr^{-1}}$, slightly surpassing the upper limit of the estimation ($\sim 3.8_{-3.1}^{+4.0} \times 10^{5}\ {\rm Gpc^{-3}\ yr^{-1}}$) in \citet{Burns_2021}, given the inclusion of the recent detection of MGF GRB 231115A. This high-event-rate density, combined with temporal similarities to short GRBs, as summerized in Table \ref{tab:summary}, suggests that MGFs could constitute a subset of short GRBs. Therefore, a comprehensive analysis of the spectral properties of MGFs is essential to further identify and understand the nature of such events in existing GRB archival data.
\section{Spectral Fit}
\label{sec:spectral}
Thanks to the high temporal and spectral resolution of the Fermi/GBM data, we conducted both time-integrated and time-resolved spectral fits and implemented the Comptonized fireball model to thoroughly examine the underlying radiation components.
\subsection{Empirical Spectral Model Fit}
We first performed both time-integrated and time-resolved spectral fits by adopting two empirical spectral models, namely cutoff power law (CPL) and blackbody (BB), over the entire event period, from -18 to 61 ms. To evaluate the goodness of fit, we examine the reduced statistic PGSTAT/degrees of freedom (dof), where PGSTAT \citep{1996ASPC..101...17A} is employed as the likelihood for Poisson data with Gaussian background, and dof is the degree of freedom. We conduct the model comparison on the basis of Bayesian information criterion \citep[BIC;][]{bic_ref}. The best-fit parameters obtained for each model within different time intervals are listed in Table \ref{tab:specfit}.

It is noteworthy that the best-fit low-energy photon index, $\alpha$, significantly surpasses the synchrotron ``death line" defined by $\alpha = -2/3$ \citep{Preece_1998} and is generally above zero, indicating that the spectra are thermal-like. An intensity tracking pattern \citep{1983Natur.306..451G} emerges in the behavior of the peak energy and the temperature throughout this event, as depicted in the left panel of Figure \ref{fig:spec_evo}. The rapid evolution of $E_{\rm p}$ and temperature in the initial two time slices indicates a sudden variation in the emission source. Subsequent decay evident in the latter two time slices may signify a cooling phase within the emission region. Such spectral evolution aligns with the characteristics expected of a rapidly expanding, followed by gradually cooling fireball-like emission source, prompting us to perform a physical fit involving a Comptonized fireball model to explain the observation. This further investigation aims to understand the origin and radiation mechanism of the event in detail.
\subsection{A Physical Model Fit}
\subsubsection{The Comptonized Fireball Model}
Consider a trapped fireball bubble, formed by photon-rich pair plasma captured by closed magnetic field lines, breaking free from its magnetic constraints and undergoing expansion toward the photosphere radius. Concurrently, due to the pressure from photon-pair plasma and the acceleration caused by the gap potential difference, a substantial number of $e^\pm$ pairs--characterized by high density and a thermal distribution--propagate along the magnetic field lines, forming a relativistic wind. Consequently, within the magnetar wind region, these relativistic $e^\pm$ pairs Comptonize the photons of the fireball, resulting in the production of high-energy gamma-ray emissions with a thermal-like distribution.

Under a strong magnetic field condition, the thermal photons of an expanding fireball undergo two fundamental scattering processes: coherent Compton (CC) and incoherent inverse Compton (IC) scattering. Hence, the Comptonized fireball model anticipates a modified thermal-like spectrum characterized by three components, each predominantly influenced by the Rayleigh-Jeans regime, coherent Compton scattering and inverse Compton process, from the low-energy end to the high-energy tail. According to \citet{10.1093/mnras/stad443}, the observed flux can be written in the form of
\begin{equation}
\label{equ:fnu_obs_zhao}
F_{\rm \nu_{obs}} = F_{\rm \nu_{obs}}^{\rm CC}(n_{\pm}, kT^\prime, B_{*}, l_0, \langle\theta_B\rangle) + F_{\rm \nu_{obs}}^{\rm IC}(kT^\prime, \alpha_{\rm IC}, l_0),
\end{equation}
where $n_{\pm}$, $T^\prime$ and $\alpha_{\rm IC}$ are the number density of the $e^{\pm}$ in the emission region, the thermodynamic equilibrium temperature in the comoving reference frame, and the index related to the IC intensity, respectively. $l_0$ denotes the initial radius of the expanding fireball. $B_{*}$ stands for the local surface magnetic field of the magnetar, which is assumed to be constant across different time intervals.

In Eq. \ref{equ:fnu_obs_zhao}, the parameter $\langle\theta_B\rangle$ is considered as an effective incident angle between the photons and the magnetic field, serving as an average effect and consolidating the impact of all possible incident angles $\theta_B$ of each individual photon. As noted in \citet{10.1093/mnras/stad443}, information on the actual values of $\theta_B$ as well as their distributions is limited and may only be viable through numerical simulation. Thus, \citet{10.1093/mnras/stad443} treat $\langle\theta_B\rangle$ as a free parameter and have attempted to obtain some constraints from spectral fitting. However, as shown in \citet{10.1093/mnras/stad443}, $\langle\theta_B\rangle$ is loosely constrained by comparing the model to data, motivating us to further investigate the flux dependence of $\langle\theta_B\rangle$.

Indeed, upon checking Eq. (17) in \citet{10.1093/mnras/stad443}, the only factor involving $\theta_B$ is $f(\theta_B) = (1 + \sqrt{\frac{\kappa_{\rm ff} + \kappa_{\rm es}}{\kappa_{\rm ff}}})^{-1}$ (note that $\kappa_{\rm ff}$ is the linear superposition of the absorption factor for O-mode and E-mode photons of the bremsstrahlung process, and $\kappa_{\rm es}$ is the superimposed scattering opacity from $e^\pm$ plasma. They are both functions of photon energy $E$). In the left panel of Figure \ref{fig:factor}, we plot $f$ as a function of $\theta_B$ and $E$ and find that, for a certain energy $E$, $f$ is almost constant for different values of $\theta_B$ (right panel of Figure \ref{fig:factor}). Thus, for a bunch of photons, the effective $\langle\theta_B\rangle$ is not sensitive to their distribution form. For simplicity, in this work, we assumed the photon incident angles, $\theta_B$, are isotropically distributed between 0 and $\pi$. We then further calculate the averaged $f$ value by averaging $f(\theta_B,E)$ over those angles, i.e.,
\begin{align}
\langle f(\theta_B,E)\rangle =
&\frac{\int f(\theta_B,E)d\Omega}{\int d\Omega} = \frac{\int_{0}^{\pi}f(\theta_B, E)\sin\theta_B d\theta_B}{2}. 
\
\end{align}

Replacing $f(\theta_B,E)$ with $\langle f(\theta_B,E)\rangle$ in Eq.\ref{equ:fnu_obs_zhao}, we finally obtain
\begin{equation}
\label{equ:fnu_obs_yh}
F_{\rm \nu_{obs}} = F_{\rm \nu_{obs}}^{\rm CC}(n_{\pm}, kT^\prime, B_{*}, l_0) + F_{\rm \nu_{obs}}^{\rm IC}(kT^\prime, \alpha_{\rm IC}, l_0),
\end{equation}
which can be used to directly fit to the observed data in \S \ref{sec:fit}.
\begin{figure*}
 \centering
 \includegraphics[width = 0.90\textwidth]{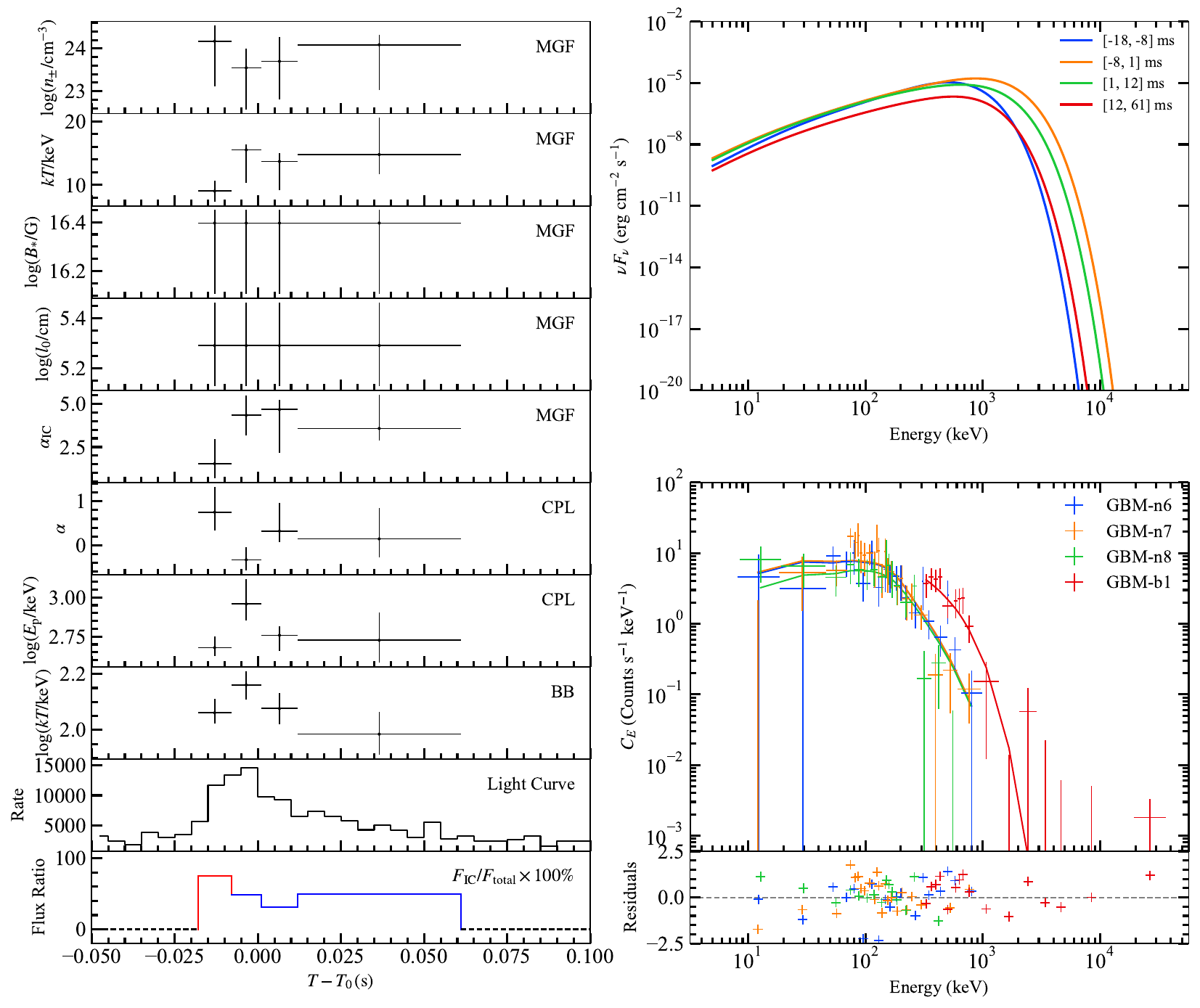}
\caption{{\it Left panels}: the observed light curve of MGF GRB 231115A and its spectral evolution based on the best-fit parameters of the Comptonized fireball model, CPL, and BB models, along with the derived flux ratio between the IC process-dominated Wien spectrum and the entire spectrum in the energy range of 1-10000 keV. The red curve represents IC domination with an IC flux ratio surpassing 50\%, while the blue curve represents CC domination with an IC flux ratio below 50\%. {\it Right top panel}: the evolution of the $\nu F_\nu$ spectra as a function of the observed times corresponding to the time-resolved slices listed in Table \ref{tab:specfit}. {\it Right bottom panels}: the observed and modeled photon count spectra of the time-integrated slice. All error bars mark the 1 $\sigma$ confidence level.}
 \label{fig:spec_evo}
\end{figure*}
\subsubsection{The Fit}
\label{sec:fit}
The fitting is performed by utilizing the Python package, \textit{MySpecFit}, following the methodology outlined in \citet{2022Natur.612..232Y,2023ApJ...947L..11Y}. The modified Comptonized fireball model with linked parameters $B_{*}$ and $l_0$ between all time slices was effectively employed to fit the observed time-resolved spectra of MGF GRB 231115A. We also performed a time-integrated spectral fit by fixing $B_{*}$ and $l_0$ to those values obtained from the time-resolved spectra fits. For all the fits, the prior ranges of the free parameters are listed in Table \ref{tab:prior}.

\begin{figure*}
 \centering
 \begin{tikzpicture}
 \node (image) at (0,0) {\includegraphics[width=0.90\textwidth]{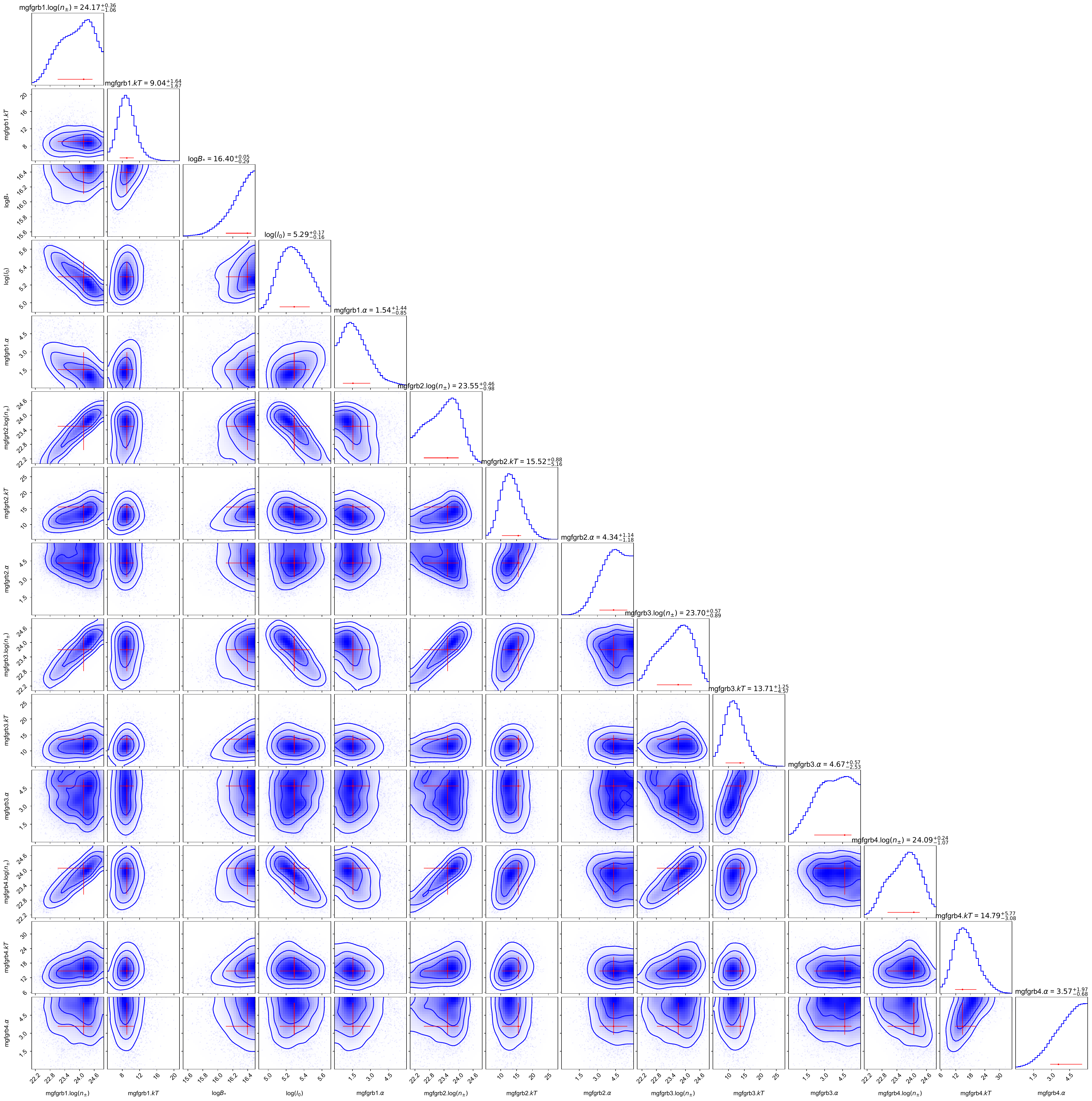}};
 \node [anchor=north east] at (image.north east) {\includegraphics[width=0.4\textwidth]{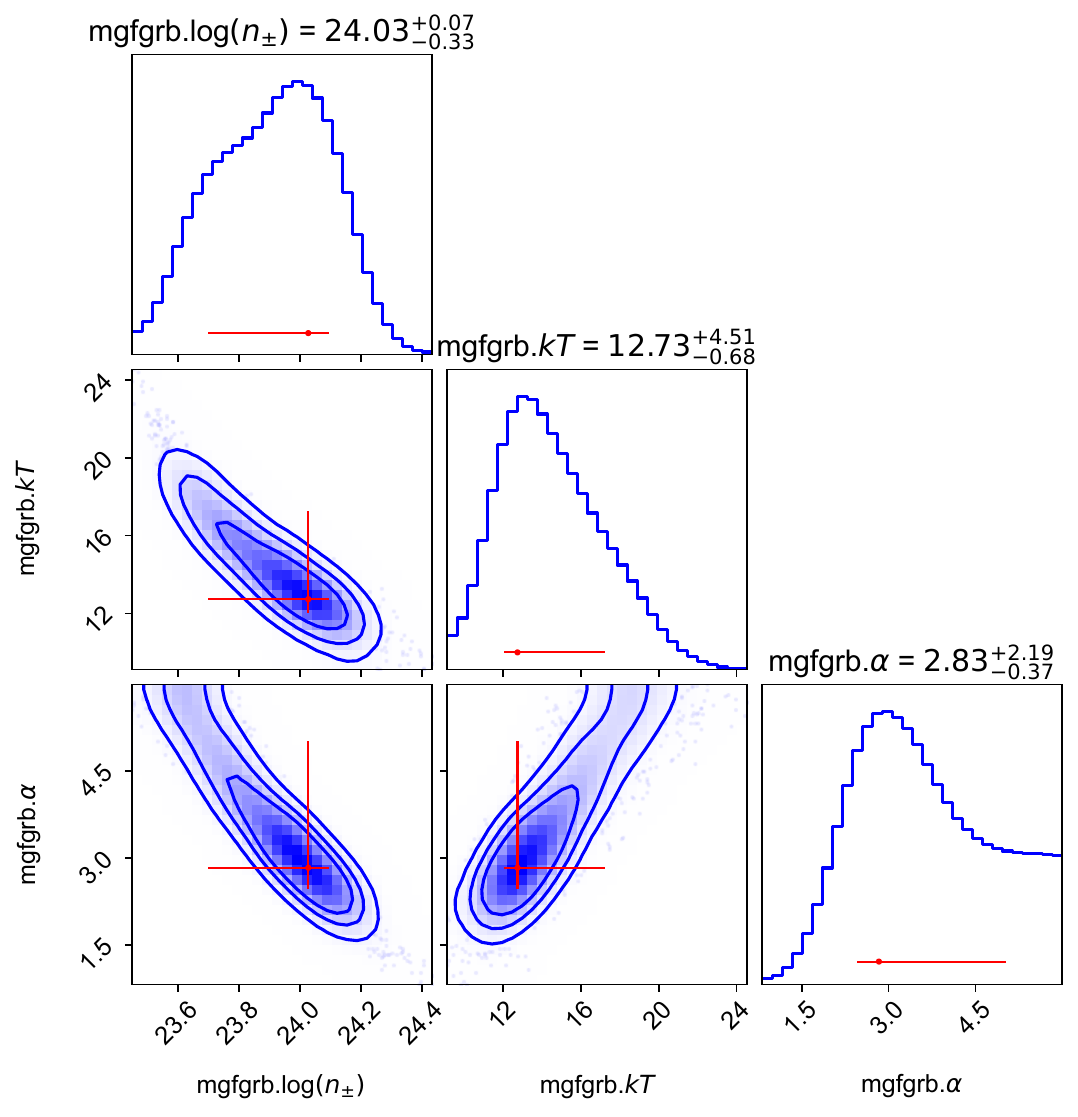}};
 \end{tikzpicture}
 \caption{{\it Left bottom panel}: corner plot of the posterior probability distributions of the parameters for the linked fit of time-resolved spectra using the modified Comptonized fireball model. {\it Right upper panel}: corner plot of the posterior probability distributions of the parameters for the fit of time-integrated spectrum using the modified Comptonized fireball model with fixed $B_{*}$ and $l_0$. The red error bars represent 1 $\sigma$ uncertainties.}
 \label{corner}
\end{figure*}

We derived best-fit parameter sets, along with their associated uncertainties (see also Figure \ref{fig:spec_evo}), and corresponding statistics detailed in Table \ref{tab:specfit}. The corner plots of the posterior probability distributions of the parameters for the fit are shown in Figure \ref{corner}. The left panel of Figure \ref{fig:spec_evo} displays the evolution of the best-fit parameters. Based on those fits, we present the evolution of the $\nu F_{\nu}$ spectra for different observed times, alongside the comparison between the observed and modeled photon count spectra for the time-integrated slice in the left panels of Figure \ref{fig:spec_evo}.
\begin{table}
\normalsize
\centering
\caption{The Prior Ranges for the Free Parameters of the Comptonized Fireball Model}
\label{tab:prior}
\setlength{\tabcolsep}{8mm}{
\begin{tabular}{lc}
\hline
\hline
Parameters & Prior \\
\hline
log$(n_{\pm})$ & [22.0, 25.0] \\
$kT^\prime$ & [0.01, 100.0] \\
log$(B_{*})$ & [14.5, 16.5] \\
log$(l_0)$ & [3.0, 6.0] \\
$\alpha$ & [0.0, 6.0] \\
\hline
\hline
\end{tabular}}
\end{table}
\begin{figure}
 \centering
 \includegraphics[width = 0.45\textwidth]{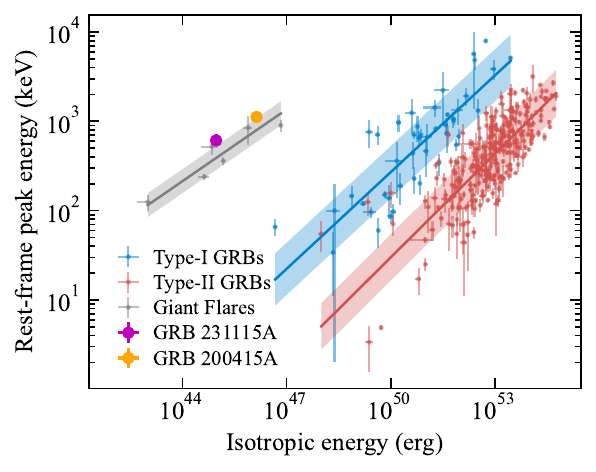}
 \caption{The $E_{\rm pz}-E_{\rm iso}$ correlation diagram. The red, blue, and gray solid lines represent the best-fit correlations for Type-I, Type-II, and MGF populations, respectively. The yellow dot marks the position of GRB 200415A from \citet{Yang_2020}. The purple dot marks the position of MGF GRB 231115A. All error bars on data points represent their 1 $\sigma$ confidence level.}
 \label{fig:epeiso}
\end{figure}
\subsubsection{The Results and Implications}
Examining the PGSTAT/dof values in Table \ref{tab:specfit}, both empirical models and the Comptonized fireball model achieve good fits. Notably, the statistical preference for the Comptonized fireball model, indicated by the smallest BIC across all time slices, emphasizes its effectiveness. This suggests the robustness of our physical model in providing a more comprehensive and adequate description of the observed data in terms of understanding the underlying radiation mechanism. Consequently, it further confirms the MGF origin of GRB 231115A.
\begin{table*}
\centering
\caption{Time-integrated and Time-resolved Spectral Fitting of GRB 231115A}
\label{tab:specfit}
\begin{tabular}{lcccccccc}
\hline
\hline
Time Intervals & \multicolumn{4}{c}{CPL Parameters} & & \multicolumn{3}{c}{BB Parameters} \\
\cline{2-5}
\cline{7-9}
($t_1$, $t_2$) (s) & $\alpha$ & $E_{\rm p}$ (keV) & PGSTAT/dof & BIC & & $kT$ (keV) & PGSTAT/dof & BIC \\
\hline
(-0.018, 0.061) & ${0.16}_{-0.19}^{+0.21}$ & ${605.54}_{-67.84}^{+84.72}$ & 468.89/463 & 487.32 & & ${119.66}_{-8.12}^{+8.32}$ & 487.51/464 & 499.79 \\
\hline
(-0.018, -0.008) & ${0.75}_{-0.42}^{+0.58}$ & ${478.50}_{-54.65}^{+84.43}$ & 321.99/463 & 340.42 & & ${115.16}_{-9.85}^{+13.69}$ & 322.62/464 & 334.91 \\
(-0.008, 0.001) & ${-0.33}_{-0.24}^{+0.29}$ & ${907.55}_{-193.04}^{+404.02}$ & 336.11/463 & 354.54 & & ${144.77}_{-16.32}^{+17.53}$ & 349.98/464 & 362.27 \\
(0.001, 0.012) & ${0.32}_{-0.24}^{+0.64}$ & ${573.73}_{-117.56}^{+83.60}$ & 313.13/463 & 331.56 & & ${119.28}_{-14.31}^{+16.46}$ & 316.62/464 & 328.91 \\
(0.012, 0.061) & ${0.14}_{-0.42}^{+0.71}$ & ${533.27}_{-149.56}^{+271.80}$ & 387.60/463 & 406.03 & & ${96.61}_{-15.14}^{+19.32}$ & 391.03/464 & 403.32 \\
\hline
\hline
 & & & & & & & & \\
\hline
\hline
Time Intervals & \multicolumn{8}{c}{MGF Parameters} \\
\cline{2-9}
($t_1$,$t_2$) (s) & log$(n_{\pm})$ & $kT^\prime$ (keV) & log$(B_{*})$ & log$(l_0)$ & & $\alpha_{\rm IC}$ & PGSTAT/dof & BIC\\
\hline
(-0.018, 0.061) & ${24.03}_{-0.33}^{+0.07}$ & ${12.73}_{-0.68}^{+4.51}$ & ${16.40}$ (fixed) & ${5.29}$ (fixed) & & ${2.83}_{-0.37}^{+2.19}$ & 467.03/463 & 485.46 \\
\hline
(-0.018, -0.008) & ${24.17}_{-1.06}^{+0.36}$ & ${9.04}_{-1.67}^{+1.64}$ & \multirow{4}{*}{${16.40}_{-0.29}^{+0.05}$} & \multirow{4}{*}{${5.29}_{-0.16}^{+0.17}$} & & ${1.54}_{-0.85}^{+1.44}$ & 321.00/461 & 334.35 \\
(-0.008, 0.001) & ${23.55}_{-0.98}^{+0.46}$ & ${15.52}_{-5.16}^{+0.88}$ & & & & ${4.34}_{-1.18}^{+1.14}$ & 336.96/461 & 350.31 \\
(0.001, 0.012) & ${23.70}_{-0.89}^{+0.57}$ & ${13.71}_{-4.57}^{+1.25}$ & & & & ${4.67}_{-2.53}^{+0.57}$ & 314.79/461 & 328.13 \\
(0.012, 0.061) & ${24.09}_{-1.07}^{+0.24}$ & ${14.79}_{-3.08}^{+5.77}$ & & & & ${3.57}_{-0.68}^{+1.97}$ & 387.96/461 & 401.30 \\
\hline
\hline
\end{tabular}
\end{table*}
The values of best-fit parameters are overall consistent with the theoretical predictions as detailedly described in \citet{10.1093/mnras/stad443}. Our results highlight the following radiation properties of this MGF burst:
\begin{enumerate}
\item $n_\pm$ is confined within the range of [$3.55 \times 10^{23}$, $1.48 \times 10^{24}$] $\rm cm^{-3}$, corroborating the notably high density of charged particles in the relativistic wind attributed to the substantial $e^{\pm}$ from the magnetosphere and the generation of secondary $e^{\pm}$ pairs. The abrupt drop of $n_\pm$ from $T_0$-18 ms to $T_0+$1 ms indicates an expansion of the fireball radius, consistent with the estimation in Eq. (3) in \citet{10.1093/mnras/stad443}, considering the same $B_{*}$ across each time interval. After $T_0+$1 ms, the number density stabilizes within a generally constant value, as indicated by the 1-$\sigma$ uncertainty. The latter evolution might be caused by the increase of secondary $e^{\pm}$ pairs while interacting with the magnetic field or the injection of the relativistic wind. 
\item $l_0$, well-constrained at $1.95 \times 10^{5}$ cm as a linked parameter in all time slices, provides us with the radius of the trapped fireball.
\item Using the time-integrated spectra, we constrained the thermal energy of the expanding fireball in the comoving frame $kT^\prime$ $\sim$ 12.73 keV. By requiring the value of Rosseland mean optical depth for the E-mode photons to be unity in Eq. (5) in \citet{10.1093/mnras/stad443}, we can derive the minimal radiation radius, $l_x$, of the expanding fireball as follows:
\begin{equation}
\label{eq:lx}
    l_x=\frac{5}{4\pi^2}\frac{\tau_\perp}{\sigma_T n_\pm}(\frac{kT^\prime}{m_e c^2}\frac{B_Q}{B_R})^{-2},
\end{equation}
where $\sigma_T$ is the Thomson cross section, $m_e$ is the electron mass, $c$ is the speed of light, $B_Q$ is the quantum critical field \citep{10.1093/mnras/275.2.255}, and $B_R$ is the magnetic field at radius $R$, i.e., $B_R=B_*(R/R_S)^{-3}$. $R_S \sim 10^6\ \rm{cm}$. Eq. \ref{eq:lx} yields $l_x \sim 1.09\times 10^6\ \rm{cm}$. We can then further calculate the value of the bulk Lorentz factor for the expanding fireball as $\Gamma \sim (l_x/l_0)^{3/2} \sim 13.30$. Utilizing $\Gamma$ and $kT^\prime$, we can estimate the observed $kT_{\rm obs}=\Gamma kT^\prime$ $\sim$ 169.31 keV. This estimation aligns with the temperature derived from the averaged BB spectrum, which stands at $\sim$ 119.66 keV, as listed in Table \ref{tab:specfit}.
\item As the model requires small-scale magnetic field lines intertwining, the increase of line density could result in the local magnetic field surpassing $10^{16}$ G. The best-fit local surface magnetic field of the neutron star, $B_{*}$, yields a value of $2.51 \times 10^{16}$ G, constrained within prior ranges.
\item $\alpha_{\rm IC}$ is constrained in the range of [0, 6] in the time-integrated time slice and the first two time-resolved slices, reflecting the domination of the IC process in the high-energy spectrum. We conducted a flux ratio estimation between the IC-dominated Wien spectrum and the entire spectrum in the 1-10000 keV energy range, revealing a significant value in each time slice shown in the left panel of Figure \ref{fig:spec_evo}. Similar findings were observed in GRB 200415A, where $\alpha_{\rm IC}$ across all time slices remains below 6. However, the last two time slices lack precise constraints even if we set the prior upper limits to 10. This outcome stems from the scarcity of high-energy photons, insufficient for a comprehensive representation of the IC process. Thus, simply elevating $\alpha_{\rm IC}$ beyond 6 would not appropriately account for the last two time slices.
\end{enumerate}
The isotropic energy predicted by the model can be derived as $E_{\rm iso}=4\pi D_L^2 F_{\rm obs}/(1+z)$, where $F_{\rm obs}$ is the model fluence calculated using the best-fit parameters and the time interval. From the time-integrated $\nu F_{\nu}$ spectrum, the peak energy $E_{\rm p}$ is determined to be $610.07_{-38.62}^{+110.68}$ keV, in agreement with $E_{\rm p}$ (CPL) at $\sim$ ${605.54}_{-67.84}^{+84.72}$ keV. The $E_{\rm p}$ and $E_{\rm iso}$ trace the giant flares track on the $E_{\rm pz}-E_{\rm iso}$ diagram \citep{2002A&A...390...81A} in Figure \ref{fig:epeiso},   indicating a broader energetic range for extragalactic MGFs, alongside GRB 200415A.
\section{Summary and Discussion}
\label{sec:sum}
In this Letter, we found that the second observed extragalactic MGF GRB, 231115A, is similar to GRB 200415A with respect to both temporal and spectral properties. We further employed a physically driven model to successfully infer the radiation origin of the burst. We conducted time-integrated and time-resolved spectral fits using the Comptonized fireball model, comparing it with empirical model fits. Our results reveal that the observed temperature and peak energy derived from the physical model align well with those obtained from CPL and BB models. Notably, the fitting statistics indicate a preference for the Comptonized fireball model. Utilizing the best-fit parameters, the Comptonized fireball model predicts local surface magnetic field instability, projecting strengths reaching up to $\sim 2.51 \times 10^{16}$ G. This instability leads to strong magnetic reconnection, forming the trapped fireball with a radius of $\sim 1.95 \times 10^{5}$ cm. Meanwhile, the high-density electrons and positrons manifest into a relativistic wind, serving as a Compton cloud, causing CC and IC scattering of the photons from the fireball. Initially, the IC process gives rise to a dominant Wien spectrum component, accounting for 75.59\% of the entire flux. Subsequently, both CC and IC processes exert significant influence on the spectrum, with CC dominating the intermediate-energy region and IC prevailing in the high-energy region. Furthermore, we determine the $E_{\rm p}$ and $E_{\rm iso}$ from the physical model and place the burst onto the $E_{\rm pz}-E_{\rm iso}$ diagram, confirming its physical origin as an MGF.

The successful fit of the Comptonized fireball model to the spectra of MGF GRB 231115A establishes a clear and self-consistent scenario to explain those peculiar bursts. Additionally, our fits indicate relatively higher local magnetic fields ($2.5 \times 10^{16}$ G), increasing the likelihood of detecting gravitational waves generated by magnetar oscillations \citep{PhysRevD.83.081302}. This makes MGF GRBs promising candidates for kilohertz gravitational wave sources \citep{2022arXiv221010931T}, especially if they can occur within our Galaxy.
\begin{acknowledgments}
We are grateful to M. van Putten, Bing Zhang and Zhen-Yu Yan for the helpful discussions. We acknowledge the support from the National Key Research and Development Programs of China (2022YFF0711404, 2022SKA0130102), the National SKA Program of China (2022SKA0130100), the National Natural Science Foundation of China (grant Nos. 11833003, U2038105, U1831135 and 12121003), the science research grants from the China Manned Space Project with NO. CMS-CSST-2021-B11, the Fundamental Research Funds for the Central Universities, and the Program for Innovative Talents and Entrepreneurs in Jiangsu. This work was performed on an HPC server equipped with two Intel Xeon Gold 6248 modules at Nanjing University. We acknowledge IT support from the computer lab of the School of Astronomy and Space Science at Nanjing University.
\end{acknowledgments}

\appendix
\section{the factor \lowercase{$f$}}
Figure \ref{fig:factor} displays the three-dimensional plot of the factor $f$ as a function of energy and incident angle, and the factor $f$ as a function of incident angle with different energies.
\setcounter{figure}{0}
\renewcommand{\thefigure}{A\arabic{figure}}
\begin{figure}
 \centering
 \includegraphics[width = 0.90\textwidth]{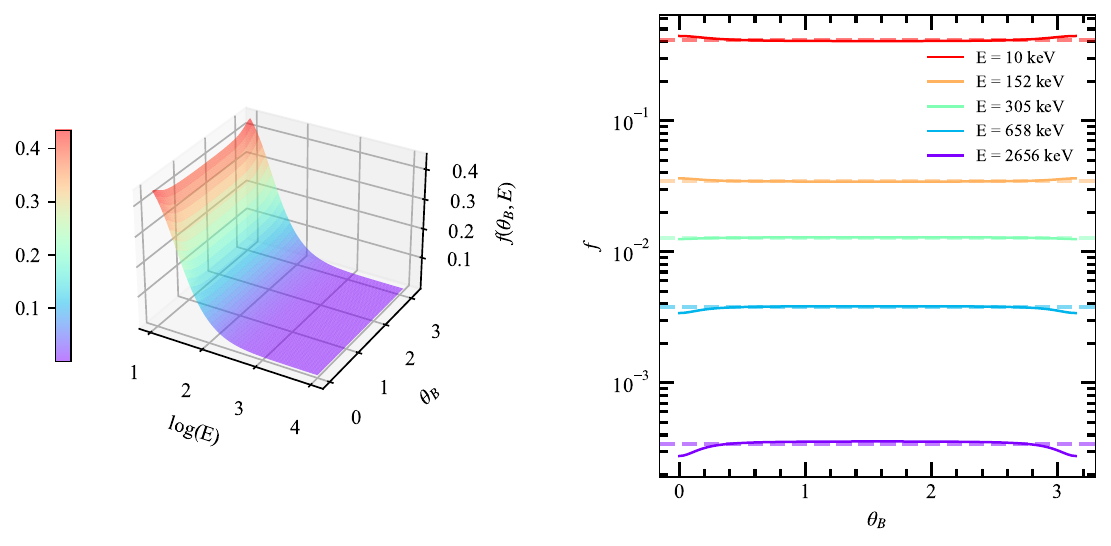}
\caption{{\it Left panel}: three-dimensional plot of the factor $f$ as a function of energy and incident angle.
{\it Right panel}: the factor $f$ as a function of incident angle with different energies. The solid curve represents the factor, while the dashed line represents the isotropically averaged factor.}
 \label{fig:factor}
\end{figure}
\end{document}